\documentclass[useAMS,usenatbib]{mn2e}
\usepackage{graphicx}
\graphicspath{{./paper_2_plots/}}

\usepackage{natbib}
\usepackage{times}
\usepackage{natbib}
\usepackage{amsmath}
\usepackage{amssymb}
\usepackage{textcomp}
\usepackage[normalem]{ulem}
\usepackage{multirow}
\usepackage{capt-of}
\usepackage{array}
\usepackage{tikz}
\usetikzlibrary{trees}
\usepackage[colorinlistoftodos]{todonotes}
\usepackage{cases}
\usepackage[export]{adjustbox}

\def \aap  {A\&A}
\def \aaps  {A\&AS}
\def \aj  {AJ}
\def \apj  {ApJ}
\def \apjs  {ApJS}
\def \araa  {ARA\&A}

\def \mnras {MNRAS}
\def \nat {Nature}
\def \apjl {ApJL}

\title[Structure, SF and environment at intermediate z]{The effect of the environment on the structure, morphology and star-formation history of intermediate-redshift galaxies}
\author[Kshitija Kelkar et al.]{Kshitija Kelkar$^{1}$\thanks{E-mail:
ppxkk1@nottingham.ac.uk (KK) },Meghan E. Gray$^{1}$, Alfonso Arag\'{o}n-Salamanca$^{1}$, \newauthor Gregory Rudnick$^{2}$, Bo Milvang-Jensen$^{3}$, Pascale Jablonka$^{4}$, Tim Schrabback$^{5}$ \\
$^{1}$School of Physics \& Astronomy, University of Nottingham, Nottingham NG7 2RD, UK\\
$^{2}$Department of Physics and Astronomy, University of Kansas, KS 66045-7582 \\
$^{3}$Dark Cosmology Centre, Niels Bohr Institute, University of Copenhagen, Juliane Maries Vej 30, 2100, Copenhagen, Denmark\\
$^{4}$Laboratoire d'Astrophysique, École Polytechnique Fédérale de Lausanne (EPFL), Observatoire de Sauverny, 1290, Versoix, Switzerland;\\
GEPI, Observatoire de Paris, CNRS UMR 8111, Université Paris Diderot, 92125, Meudon Cedex, France \\
$^{5}$Argelander Institut fuer Astronomie, Auf dem Huegel 71, D-53121 Bonn, Germany}

\begin{document}
\date{Accepted xxx. Received xxx; in original form xxx}

\pagerange{\pageref{firstpage}--\pageref{lastpage}} \pubyear{xxxx}

\maketitle

\label{firstpage}

\begin{abstract}
  With the aim of understanding the effect of the environment on the star formation history and morphological transformation of galaxies, we present a detailed analysis of the colour, morphology and internal structure of cluster and field galaxies at $0.4 \le z \le 0.8$. We use {\em HST} data for over 500 galaxies from the ESO Distant Cluster Survey (EDisCS) to quantify how the galaxies' light distribution deviate from symmetric smooth profiles. We visually inspect the galaxies' images to identify the likely causes for such deviations. We find that the residual flux fraction ($RFF$), which measures the fractional contribution to the galaxy light of the residuals left after subtracting a symmetric and smooth model, is very sensitive to the degree of structural disturbance but not the causes of such disturbance. On the other hand, the asymmetry of these residuals ($A_{\rm res}$) is
more sensitive to the causes of the disturbance, with merging galaxies having the highest values of $A_{\rm res}$. Using these quantitative parameters we find that, at a fixed morphology, cluster and field galaxies show statistically similar degrees of disturbance. However, there is a higher fraction of symmetric and passive spirals in the cluster than in the field. These galaxies have smoother light distributions than their star-forming counterparts. We also find that while almost all field and cluster S0s appear undisturbed, there is a relatively small population of star-forming S0s in clusters but not in the field. These findings are consistent with relatively gentle environmental processes acting on galaxies infalling onto clusters.

\end{abstract}


\section{Introduction}
\label{secintro}

Galaxy clusters represent an excellent agglomeration of galaxy populations undergoing changes in several observable galaxy properties, some of which are attributed to the diversity of environments that the galaxies experience. One
of the earliest suggestions that environment may play a role in
transforming galaxy properties is the well established
morphology--density relation \citep{dressler1980,dressler84}: high
density environments are observed to contain higher fractions of
galaxies with early-type morphologies than the field.  The question of
precisely to what extent, and by what physical processes, the
environment leaves an imprint on morphology as well as other
observable properties (e.g. colour, star formation, internal
structure) is still largely undetermined.

Evidence of global transformations happening over look-back time is given by the increasing fraction of spiral galaxies in clusters till $z\sim0.5$ \citep{dressler97,fasano00,desai07} and the fact that high-$z$ clusters are found to contain more star-forming
galaxies as compared to present-day clusters \citep{butcher78,poggianti06}.
In addition to the morphology--density relation, it is widely observed that the specific star-formation rate declines towards dense
local environments \citep{hashimoto98,lewis02,kauffmann04,gray04,balogh07}. Higher fractions of passive or quiescent galaxies are found in dense
environments, both in the local Universe \citep{baldry06,vandenbosch08,gavazzi10,haines13} and out to $z\sim2$
\citep{poggianti99,cooper11,sobral11,quadri12,woo13,kovac14}.

Some environmental segregation in the galaxies' properties is naturally expected: hierarchical
models of structure formation predict that the densest regions will
collapse at earlier times, forming the cores of clusters.  The cluster
galaxies at a given epoch will, therefore, be more evolved than the
average field galaxy \citep{dlucia04-1}.  Further, the decline in
global star-formation rate with redshift \citep{madau97,fergusson00}
will result in fewer star-forming field galaxies being accreted onto
clusters at later times.  However, as the clusters assemble and
evolve, the accreting galaxies are also subjected to various
interactions with other galaxies and the wider group or cluster
environment.

These physical processes will impact the galaxies in different ways,
affecting both star-formation rates and stellar distributions.  Strong
gravitational interactions such as mergers and strong tidal
interactions \citep{barnes92, barnes96}, are efficient in altering
galaxy structure as well as triggering star formation. Indeed, it has
been observed that most starbursts or galaxies with very high star formation display merger signatures, irrespective of redshift
\citep{duc97,elbaz03}. Recent studies like \citet{kartaltepe12}, however, show that extreme star-forming galaxies since $z\sim2$ are comprised of a mix regular morphology galaxies and galaxies showing early stages of interaction/ongoing mergers.
Tidal interactions or harassment lead to stripping of outer material
from the galaxy under the impact of high-speed encounters, resulting
in temporary enhancement of star formation \citep{boquien09, moore96}. 

While gravitational interactions may redistribute the stellar content
of the galaxy or trigger bursts of star formation, gaseous processes
also influence the star formation rate.  With $\sim 10\%$ of the total mass
of the cluster consisting of hot intracluster medium (ICM), infalling
galaxies may undergo loss of their cold disk gas through ram-pressure
stripping \citep{gunngott72} or hot gaseous halo through starvation
\citep{larson80}. Low-redshift observational studies have shown evidence of stripping of
the material from galaxies in cluster environments in the form of
`jellyfish' galaxies \citep{kenney04,merluzzi13,fumagalli14,jaffe15}.

Several types of transition objects have been identified that
represent populations of galaxies in the process of having their
star formation shut down.  For example, `post-starburst' or `k+a',
galaxies make up a significant fraction of intermediate to high-$z$
clusters, while being rare at $z=0$. Further, the strong correlation
between cluster velocity dispersion and `k+a' fraction suggests a
possibility of interactions with the ICM being responsible for the
eventually turning them passive \citep{poggianti09-2}, though it may
not be the dominant process for the transformation
\citep{delucia09}. Structurally, this could be related to the
transformation of star-forming spiral galaxies into lenticular
galaxies, as discussed by \citet{dressler97} and \citet{poggianti99},
further corroborated by the lack of blue lenticulars in clusters
\citep{jaffe11}. Indeed, \citet{gallazzi09} and \citet{wolf09} found a
cluster-specific population of smooth spiral galaxies with suppressed star formation
in the STAGES multiple-cluster system \citep{gray09}. Analysis of
rotation curves by \citet{bosch13} confirmed that these same objects
contain kinematically disturbed gas while remaining optically symmetric.  It
is clear that for these smooth passive spirals, gas processes such as starvation and ram-pressure stripping \citep{haines13} are shutting down star formation without
simultaneous wide-scale redistribution of their stellar material.

When attempting to understand the connections between changes in star
formation and structure, one key challenge is to identify the
\emph{cause} of structural disturbances. There have been many methods
developed for identifying and analysing specific gravitational
interactions such as mergers (which are capable of leaving prominent
signatures observable over long timescales). These approaches often
involve measuring the structural properties in galaxy images, like the
CAS \citep{conselice03} or Gini-M$_{20}$ \citep{lotz04} systems. Other
approaches include using multimode (M), intensity (I), and deviation (D)
statistics to identify galaxies which are likely mergers
\citep{freeman13}, or analysing the
residual light remaining when a smooth profile is removed
\citep{hoyos12}.  Each of these methods is found to be sensitive to
different stages or types of interaction, for example, the CAS criteria
tends to pick out all major mergers, whereas the Gini/M$_{20}$
measures both minor and major mergers \citep{lotz08,lotz10a}.
However, none of these methods are able to produce a complete and
uncontaminated sample of galaxy interactions or structural
disturbances, highlighting the complexity of quantifying galaxy
structure and interpreting it.  Furthermore, these methods give no
insight into the physical causes of any asymmetries (reflecting
internal brightness fluctuations, or evidence of external
gravitational influences), so visual interpretation of images is
invaluable. 

In this paper, we seek to explore the interconnected relationships
between galaxy morphology, star formation properties, and environment,
while introducing additional information about the irregularities in
the stellar distribution, as well as interpretations of the probable
cause of any disturbances.  We focus on galaxies in cluster and field
environments at intermediate redshifts within $0.4 < z < 0.8$, using
the ESO Distant Cluster Survey (EDisCS).  We aim to quantify galaxy
structure using quantitative analysis of galaxy images (complementing previous work on on bulge/disk decompositions by \citealt{simard09}), as well
incorporating visually determined information from galaxies. We
further study the correlations of galaxy structure with the observed
photometric properties of galaxies and global environment (this paper)
and eventually linking them to the star formation history of galaxies
and the local environment (Kelkar et al. 2017, in prep).

This paper is structured as follows: Sections~\ref{secdata} and
\ref{subsecvisdist} describe the data, the sample selection, and the
methodology used when defining the environment and defining galaxy
structure.  Sections~\ref{secgalaxystruct}, \ref{sec:VQstruct}, and
\ref{sec:structSF} analyse and discuss the galaxies'
structure, photometric properties and environment. Finally, in
Section~\ref{secconclusions} we present a discussion of our results
and conclusions. Throughout this paper, we use the standard
$\Lambda$CDM Cosmology with $h_{0} = 0.7$, $\Omega_{\Lambda} = 0.7$
and $\Omega_{\rm m} = 0.3$. When relevant, we use a Kroupa IMF \citep{kroupa01} and AB magnitudes\footnote{\textbf{The original EDisCS papers published Vega magnitudes. These were converted into the AB system by Rudnick et al.\ (2017, submitted).}}.

\section[]{Data}
\label{secdata}

The data analysed in this paper was described in detail in \cite{kelkar15}. To avoid repetition, we only provide here 
a brief summary of the most relevant information. We refer the interested reader to that paper. 

Our data originate from the ESO Distant Cluster Survey \citep[EDisCS;][]{white05}, which studied 20 fields containing galaxy 
clusters from the Las Campanas Distant Cluster Survey \citep{gonzales01} in the redshift range $0.4<z<1$. 
Optical imaging in the $V$, $R$ and $I$ bands was obtained with FORS2 on the {ESO Very Large Telescope} \citep[VLT;][]{white05}. 
Near-IR $J$ and $K_{\rm s}$ photometry from SOFI at the $3.5\,$m {New Technology Telescope (NTT)} is also available \citep{rudnick09}.
Spectroscopy with FORS2/VLT was obtained for an effectively $I$-band-selected sample of galaxies with redshifts at or near the cluster redshifts \citep{halliday04,mjensen08}.

In addition, the cluster fields studied here also have {\em HST} $I$-band ($F814W$) imaging obtained with the ACS camera \citep{desai07}. A total of five pointings were taken in each field, four adjacent one-orbit pointings covering $6.5^\prime \times 6.5^\prime$ (approximately the field of the VLT optical images) and an additional four-orbit pointing covering the central $3.3^\prime \times 3.3^\prime$ region of each cluster. Mosaic stacks that encompass all ACS tiles for a given cluster were created employing \texttt{MultiDrizzle} \citep{koekemoer03}, and scripts for optimised image registration and weighting as detailed in \citet{schrabback10}. The work presented in this paper exploits the {\em HST} imaging to carry out the structural analysis of the galaxies. Table~\ref{table1} gives a summary of the properties of the cluster sample.

Other follow-up data for these clusters include {\em Spitzer} IRAC ($3-8\mu m$) and MIPS ($24\mu m$) imaging, H$\alpha$ narrow-band imaging for three of the fields \citep{finn05}, and {\em XMM-Newton/EPIC} X-ray observations for a subset of the clusters \citep{johnson06}.    

{\em HST-}based visual galaxy morphologies were published by \citet{desai07}. For the purposes
of this study, we have collapsed the fine morphological classes given by the original catalogue into four broad bins: ellipticals,
lenticulars, spirals, and irregulars.

\begin{table}

  \caption{ Summary of the cluster sample properties (including secondary clusters identified along the line-of-sight, cf. \S\ref{subsecenv}), sorted according to cluster halo mass. Columns~1--5 contain the cluster ID,  cluster redshift, cluster velocity dispersion, cluster halo mass \citep[calculated following][]{finn05} and the number of spectroscopically confirmed cluster members \citep{halliday04,mjensen08}.  \label{table1}}
\begin{center} 
\begin{tabular}{lcccc}
 
 \hline
 Cluster &  $z_{\rm cl}$ & $\sigma_{\rm cl}$ & $\log M_{\rm cl}$ & No.\ of spec. \\
 &  & (km$\,$s$^{-1}$) &  ($M_\odot$) & members \\
 \hline
 {\it Clusters} & & & & \\
 cl1232$-$1250  & 0.5414 & 1080$^{+119}_{-89}$ & 15.21 & 54 \\
 cl1216$-$1201  & 0.7943 & 1018$^{+73}_{-77}$ & 15.06 & 67 \\
 cl1138$-$1133  & 0.4796 & 732$^{+72}_{-76}$ & 14.72 & 49 \\ 
 cl1354$-$1230   & 0.7620 & 648$^{+105}_{-110}$ & 14.48 & 22 \\
 cl1054$-$1146   & 0.6972 & 589$^{+78}_{-70}$ & 14.38 & 49 \\
 cl1227$-$1138  & 0.6357 & 574$^{+72}_{-75}$ & 14.36 & 22 \\
 cl1138$-$1133a   & 0.4548 & 542$^{+63}_{-71}$ & 14.33 & 14 \\
 cl1037$-$1243a & 0.4252 & 537$^{+46}_{-48}$ & 14.33 & 43 \\
 cl1054$-$1245   & 0.7498 & 504$^{+113}_{-65}$ & 14.16 & 36 \\
 cl1040$-$1155  & 0.7043 & 418$^{+55}_{-46}$ & 13.93 & 30 \\
 cl1227$-$1138a   & 0.5826 & 432$^{+225}_{-81}$ & 13.69 & 11 \\
 {\it Groups} & & & & \\
 cl1103$-$1245a  & 0.6261 & 336$^{+36}_{-40}$ & 13.66 & 15 \\
 cl1037$-$1243 & 0.5783 & 319$^{+53}_{-52}$ & 13.61 & 16 \\
 cl1103$-$1245b  & 0.7031 & 252$^{+65}_{-85}$ & 13.27 & 11 \\
 
 \hline

\end{tabular}
\end{center}
\end{table}

\subsection[]{Environment definition}
\label{subsecenv}

We separate the sample by global environment based on spectroscopic cluster membership. A galaxy is considered a member of a cluster if its spectroscopic redshift lies within $\pm3\sigma_{\rm cl}$ from the average cluster redshift $z_{\rm cl}$ \citep{mjensen08,halliday04}. All the galaxies that do not satisfy this criterion are considered to be in the field sample. Although the redshift distribution of cluster and field galaxies are very similar, to avoid potential biases we only consider field galaxies whose redshifts are contained within the redshift range spanned by the clusters (with a $z$ tolerance of $\pm0.05$ at each end, i.e., from the lowest and the highest cluster redshift in our sample).

Some of the EDisCS fields contain secondary clusters in addition to the main ones \citep{white05, mjensen08}. Members of these secondary clusters are, for consistency, also included in the cluster  sample. These secondary clusters are denoted in Table~\ref{table1} with `a' or `b' following the main cluster ID. \citet{poggianti09-2} classified these secondary structures into clusters and groups. Structures with $\sigma_{\rm cl} > 400$ km$\,$s$^{-1}$ were classed as `clusters', while structures with $160 < \sigma_{\rm cl} < 400$ km$\,$s$^{-1}$ and at least 8 spectroscopic members were classed as `groups'. In this paper, the global environment of the galaxies is defined based on their cluster membership irrespective of the host cluster/group identification.

\subsection[]{Sample selection}
\label{subsecsample}

In what follows, we will use both the whole spectroscopic sample defined in Section~\ref{subsecenv} (to maximise the number of galaxies) and a mass-complete subsample containing 265 galaxies with a stellar mass completeness limit of log $M_{*}/M_{\odot}$=10.6 \citep{vulcani10}. The mass-complete subsample will be used to ensure that no mass-related biases affect our conclusions.

Note that both samples contain only galaxies whose spectra have a $S/N$ ratio in the continuum that is larger than $2$. This ensures not only the reliability of the redshifts, but also a reasonable quality in the measurements of spectral features such as the $4000$\AA\ break, the [OII]$\lambda3727$ emission line, and several strong Balmer absorption lines. These spectral features are analysed in Kelkar et al. (2017, in prep) and Rudnick et al.\ (2017, submitted) using similarly defined samples for direct comparison. Table~\ref{samptbl} provides some information on these samples.

\begin{table}
\caption{\label{samptbl} Details of the full spectroscopic sample and subsample, divided by environment and morphology. The subsample has a stellar mass-completeness of log $M_{*}/M_{\odot}=$ 10.6. \newline}
 \centering
 \begin{tabular}{lrrrrrr}
\hline
\multicolumn{2}{l}{\bf Spectroscopic sample}              & E   & S0 & Sp  & Irr & Total \\
\hline
Cluster & All           & 104    & 46   & 195    & 16    & 361 \\
 & Mass-complete &  65   & 30   &  95   &  4   & 194 \\
\hline
Field & All           &  31   &  9  &  91   & 20    & 151 \\
 & Mass-complete &  15   & 6   & 35    &  1   &  57\\
\hline

\hline
\end{tabular}
\end{table}

\section[]{Visual classification of structural disturbances}
\label{subsecvisdist}

To complement the information provided by the galaxies' morphological Hubble types \citep{desai07}, in this paper we qualitatively analyse galaxy `structure' by studying the detailed distribution of galaxy light. For this purpose, we use the terms `asymmetry' to refer to visible departures from a symmetric light profile, and `disturbance' to indicate our assessment of whether the cause of that deviation is internal or external in origin. Therefore, a galaxy may have a combination of `asymmetry' and `disturbance' properties. For instance, a galaxy may be symmetric and undisturbed; another may be internally asymmetric but undisturbed (e.g. an otherwise symmetric spiral galaxy with a prominent HII region); a third one may be asymmetric due to an external distortion (e.g., gravitational interaction). To clarify all these possible categories, Figure~\ref{flowchart} gives a graphical representation of the classification scheme, described below.

This classification was carried out by visually inspecting {\em HST} images of all the galaxies in our sample taken in the $I$-band (corresponding, approximately, to the rest-frame $B$-band). Three of the authors (KK, AAS, MEG) performed independent classifications of every galaxy. Note that these classifications were carried out blindly, without knowledge of the cluster membership of the galaxies, their redshifts or their morphology type.  

\subsection{Visual asymmetry class} 

As a first step in the classification, we separate galaxies into two distinct classes, `symmetric' and `asymmetric'. This is done by visually identifying asymmetric features in the galaxies’ images as possible indicators of structural disturbances. Explicitly, we classified galaxies as `asymmetric' if they possess asymmetric features, and `symmetric' in the absence of them.

\subsection{Visual disturbance class} 

\begin{figure*} 
  \centering

  \includegraphics[width=1.1\textwidth]{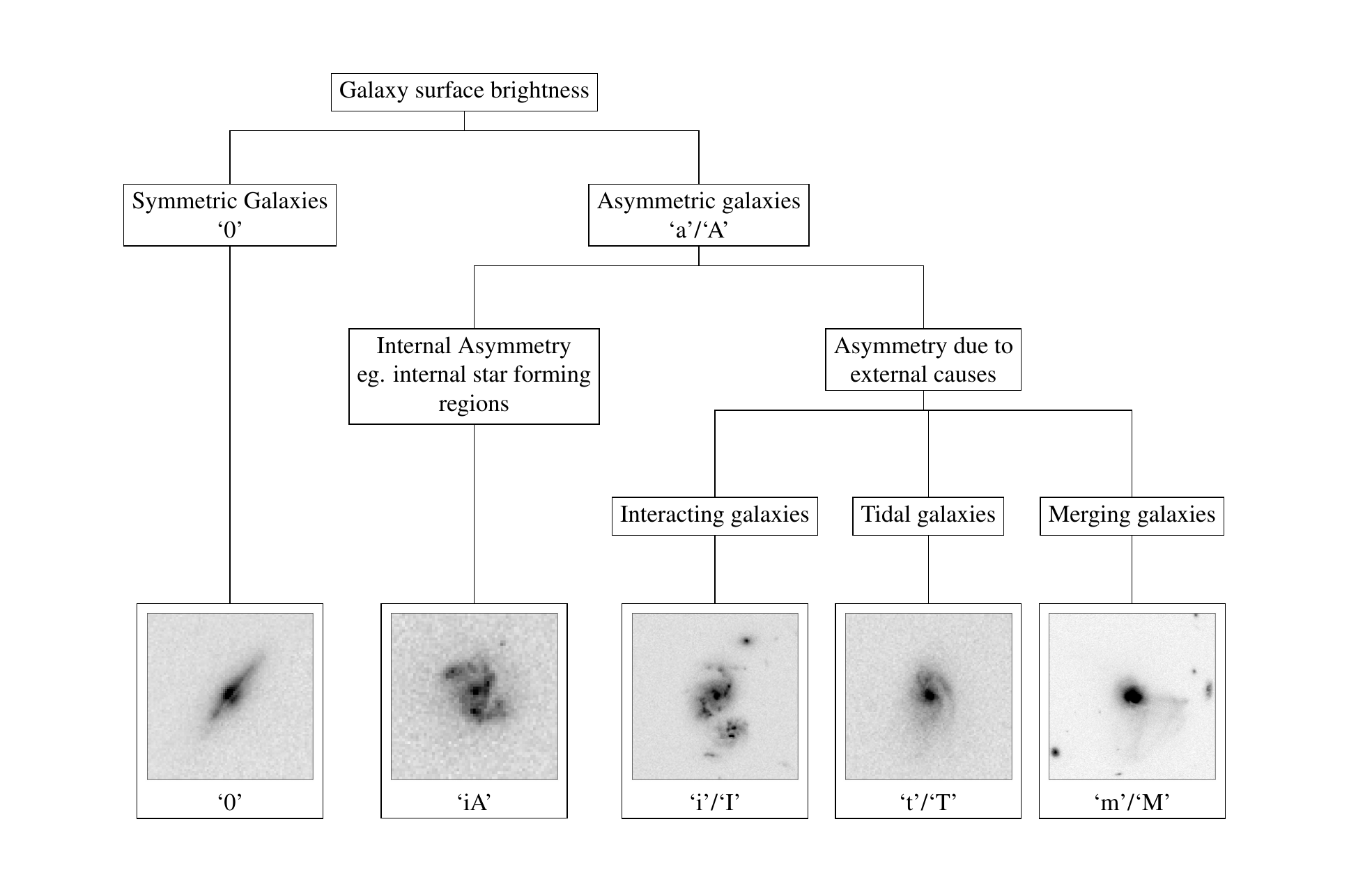}
   \caption[Graphical representation of visual classification scheme for identifying structural disturbances]{\label{flowchart} Graphical representation of our
   classification scheme for identifying structural disturbances. We
   use a two-stage process: first visually identifying galaxies with
   some form of asymmetry in their surface brightness distribution,
   then further refining that classification based on an evaluation of
   the probable cause of the disturbance (whether internal or external
   in origin).  The images are {\em HST} thumbnails of representative
   galaxies from the sample identified in each class. Additionally, a
   small number galaxies classified as `Chaotic' (C) or `Undefined'
   (X), and are not included in this diagram. Please refer to
   Section~\ref{subsecvisdist} for details regarding classification scheme.}
   \end{figure*}

For those galaxies with
visual asymmetry, we further designed a \
classification scheme, independent of morphological type, to identify the probable cause of the
disturbance. Figure \ref{flowchart} gives a graphical representation of the classification
scheme, described below.
\begin{itemize}

\item \textbf{Internal Asymmetry (iA)} : The galaxies classified under
  this category showed distinct visual asymmetry due to features like
  prominent star-forming regions/knots in the galaxy. Further, these
  asymmetries showed no clear evidence of any form of external
  processes which may be acting on the target galaxy. These galaxies
  are assigned a non-zero asymmetry but no disturbance index and constitute 
  only $\sim$ 7\% of the total sample. However,
  note that such internal asymmetries may well still be the result of
  external causes like mergers \citep{bournaud08} or ram-pressure stripping events \citep{poggianti16}, even though these may not be apparent.

\item \textbf{Galaxy interaction (i/I)} : Galaxies in this class
  showed features indicating interactions with a companion galaxy. Lowercase ``i'' denotes `weak interaction', while uppercase ``I'' indicates `strong interaction', as judged by the classifier.

\item \textbf{Tidal interaction (t/T)} : Galaxies in this class
  displayed tidal features (e.g. a tail of stripped material extending
  outside the galaxy) but with no obvious sign of an interacting
  neighbour. As before, lowercase/uppercase letters are used to indicate the strength of the features.

\item \textbf{Mergers (m/M)} : We identified ongoing galaxy mergers
  based on distinct asymmetric merging signatures. Minor or major
  mergers were identified through a visual assessment of the stellar mass
  ratios involved.  Galaxies appearing as a single distorted merger
  remnant or possessing clear galaxy cores of similar brightness were
  classified as major mergers (M). Galaxies seen merging with a
  smaller galaxy were identified as minor mergers (m).  Our
  classifications are informed by the visual appearance of merging
  galaxies in simulations, and experience of classifying mergers in
  STAGES \citep{gray09} and EDisCS.

\item[$\bullet$] \textbf{Chaotic/Undefined systems (C/X)} : The final
  class contained a small number of galaxies (less than $1\%$ of the
  total sample) displaying structures that were chaotic or could not be
  associated with any of the categories above.
\end{itemize}

\subsection*{Final classifications}
\label{ssec:final_class}
Symmetric galaxies were assigned an index of `0', whereas asymmetric
galaxies were indexed as `a'/`A', with the lower/upper case of the
index denoting an assessment of the strength (weak/strong) of the
asymmetric features.  Asymmetric galaxies were then assigned a
disturbance class label according to the schema described above, with
the lower/upper case of the index denoting the mild/strong nature of
the external features. To determine the strength of the visual classification (ie weak/strong), individual indices of `0' were given a weight of 0 while lower/upper case indices were given a weight
of `1' and `2' respectively.  

Since three independent classifiers classified each galaxy, the final combined classification for asymmetry and disturbance was
determined by majority vote (independent of the index case). The final classification was then associated 
with the summed weights of the contributing indices. If all
the three classifiers disagreed, the final classification was selected
at random from the three votes. The strength of classification in this case would be the weight of the randomly selected classification index. In a
small number of cases where a classifier was not confident in the
assessment, the individual contribution was downweighted to 0.5.

Also note one further subtlety in our classification scheme. A subset of
galaxies with smooth early-type morphologies (asymmetry = `0')
nevertheless were identified on the balance of probability as having a
disturbance class (minor interaction, `i') based on the presence of a
very close neighbour. These `0$\&$i' galaxies represent possible dry
merger candidates, where a merger may be ongoing but the visual
signatures are short-lived due to an absence of gas in the galaxies \citep{bell06}.

Table \ref{table} gives the
fractions of galaxies in each disturbance class in the cluster and
field environment with respect to the entire sample. We use the \citet{wilson27} binomial confidence
interval to compute the $1\sigma$ uncertainty in the fractions
$\delta{f_{i}}$ 
\begin{equation}
\label{Equation: Frequency error}
\acute{f}_{i} \pm \delta{\acute{f}_{i}} = \frac {N_{i} + \kappa^2/{2}}{N_{\rm tot} + \kappa^2} \pm 
\frac{\kappa\sqrt{N_{\rm tot}}}{N_{\rm tot} + \kappa^2}\sqrt{f_{i}(1- f_{i}) + \frac{\kappa^2}{4N_{\rm tot}}},
\end{equation}
where $f_{i} = N_{i}/N_{\rm tot}$, and $\kappa$ is the $100(1-\alpha/2)\rm{th}$ percentile of a standard normal distribution ($\alpha$ being
the error percentile corresponding to the $1\sigma$ level (refer also to \citealt{brown01}). These fractions
will be discussed in Section~\ref{sec:structSF}. Note that even if $ N_{i}=0$, the estimated value of $\acute{f}_{i}$ is not necessarily $0$.

  \begin{table*}
    \begin{center}
    \renewcommand{\arraystretch}{1.5}
    \centering
    \caption[The relative fractions for galaxies in the mass-complete sample identified in each disturbance class, for a fixed morphology in cluster and field environment]{\label{table} The relative fractions for galaxies 
    identified as undisturbed, internally asymmetric (possessing asymmetry but with no obvious external cause), interacting galaxies, 
    tidal galaxies and galaxies experiencing an ongoing merger, for a fixed morphology in cluster and field environment. Refer to Section~\ref{sec:structSF} for detailed discussion.\newline}
    \begin{tabular}{lcccccc}

    \centering
    \textbf{Morphology} & \textbf{Environment} & Undisturbed & Internally asymmetric & Interacting & Tidal & Merger \\
  
     & & (0) & (iA) & (i/I) & (t/T) & (m/M) \\
    \hline
    \hline
    \multirow{2}{20pt}{Ellipticals (E)} & Cluster & 0.79$\pm$0.04 & 0.01$\pm$0.01 & 0.17$\pm$0.04 & 0.02$\pm$0.01 & 0.02$\pm$0.01 \\ 
    & Field & 0.64$\pm$0.08 & 0.08$\pm$0.05 & 0.14$\pm$0.06 & 0.11$\pm$0.05 & 0.08$\pm$0.05 \\ 
    \hline
    \multirow{2}{20pt}{Lenticulars (S0)} & Cluster & 0.93$\pm$0.04 & 0.01$\pm$0.01 & 0.07$\pm$0.04 & 0.01$\pm$0.01 & 0.01$\pm$0.01 \\  
    & Field & 0.85$\pm$0.11 & 0.05$\pm$0.05 & 0.15$\pm$0.11 & 0.05$\pm$0.05 & 0.05$\pm$0.05 \\
    \hline
    \multirow{2}{20pt}{Spirals (Sp)} & Cluster & 0.35$\pm$0.03 & 0.25$\pm$0.03 & 0.18$\pm$0.03 & 0.09$\pm$0.02 & 0.11$\pm$0.02 \\ 
    & Field &  0.36$\pm$0.05 & 0.22$\pm$0.04 & 0.20$\pm$0.04 & 0.11$\pm$0.03 & 0.11$\pm$0.03 \\
    \hline 
    \multirow{2}{20pt}{Irregulars (Irr)} & Cluster & 0.03$\pm$0.03 & 0.38$\pm$0.12 & 0.15$\pm$0.08 & 0.09$\pm$0.06 & 0.32$\pm$0.11 \\ 
    & Field & 0.02$\pm$0.02 & 0.21$\pm$0.09 & 0.12$\pm$0.07 & 0.21$\pm$0.09 & 0.40$\pm$0.11 \\
    \hline 
  
    \end{tabular}
    \end{center}
    \end{table*}

\section[]{Quantitative structure}
\label{secgalaxystruct}

In addition to our \emph{qualitative} assessment of galaxy asymmetry
and disturbance, we also further explore \emph{quantitative}
measurements of galaxy structure.  Specifically, we extract structural
information from the galaxy residuals after a smooth surface
brightness profile is removed. Although originally intending to
identify minor mergers, \citet{hoyos12} show that measuring structural
parameters of residuals of galaxies is a good way of determining
disturbances in galaxy structure that are otherwise faint to detect
visually but are observable over a longer timescale.

\subsection[]{Constructing galaxy residual images}
\label{resimg}

The galaxy residual images required for this analysis were obtained
using the data pipeline {\sc galapagos} \citep[Galaxy Analaysis over Large
Area: Parameter Assesement by {\sc galfit}ing Objects from {\sc
  sextractor;}][]{barden12}. All galaxies from the ten {\em HST}
$I$-band mosaics were detected using {\sc sextractor}, and
corresponding image stamps were created by {\sc galapagos}. These image
stamps were fitted with a 2D S\'{e}rsic light profile
\citep{sersic1968} using {\sc galfit} \citep{peng02,peng10}, which
resulted in generation of galaxy residual stamp images after the
S\'{e}rsic model was subtracted.  \citet{kelkar15} contains further
details of the fitting method. These residual images were used to
compute quantitative `Asymmetry' ($A_{\rm res}$) and a measurement of
the signal remaining after subtracting the S\'{e}rsic model (residual
flux fraction, `RFF').

\subsection[]{Asymmetry in residual images ($A_{\rm res}$)}
\label{qs-ares}
Using the CAS (Concentration, Asymmetry, ClumpineSs;
\citealt{bershady00}) system, we define the asymmetry `$A_{\rm res}$' in
the galaxy residual image to measure the extent of residual light under a
180 degree rotation around a point that minimizes the asymmetry of the
galaxy image. It is defined as:
  \begin{equation}
   A_{\rm res} = \left(\frac{\sum_{i,j}|I_{i,j}-I_{i,j}^{180}|}{\sum_{i,j}|I_{i,j}|}\right)-\left(\frac{\sum_{i,j}|B_{i,j}-B_{i,j}^{180}|}{\sum_{i,j}|I_{i,j}|}\right),
  \end{equation}
  Here $I_{i,j}$ represents the flux at pixel ($i$,$j$) in the galaxy residual
  image whereas $I_{i,j}^{180}$ represents the same image rotated through
  180 degrees. The second term in the equation accounts for the
  background contribution. We construct a background noise image to
  compute the second term using the EDiSCS noise images for the {\em
    HST} ACS mosaics. As with the construction of the residual images,
  associated noise images were cut out for individuals galaxies with the
  same dimensions as the residual images. 

  As a first step, these noise stamp images were multiplied with the
  exposure time corresponding to the region in the mosaic (Refer to
  \S\ref{secdata}). This modified image was then multiplied by a
  white noise image with $\sigma = 1$. The resultant image is a good
  representation of the background noise. Both the terms in the above
  equation are computed over an aperture defined by constructing an
  ellipse whose semimajor axis is the radius of Kron
  aperture\footnote{ In this paper, we use the definition of
    radius of the ``Kron aperture'': 2.2$r_{1}$, where $r_{1}$ is the
    first moment of the light distribution \citep{bertin96}. This
    corresponds to the semimajor axis for an elliptical light
    distribution.} and are minimised independently. We implement a
  slightly modified method for minimising these terms, deviating from
  the original recipe described in \citet{conselice00}. We allow the
  centre of rotation to lie at a maximum of 3 pixels in radius from
  the {\sc sextractor} defined centre over a grid of predefined points
  1 pixel apart. The main advantage of this new method is that the pixel values are not interpolated under 180-degree rotation due to the choice of integral rotation centres. Moreover, one could think of this method computing global rotational asymmetry, and hence reducing the computation time as compared to the original method. The possible values that $A_{\rm res}$ can take ranges from 0 to 2. 

\subsection[]{Residual flux fraction ($RFF$)}
\label{qs-rff}

The second quantitative diagnostic we use is the Residual Flux fraction \citep{hoyos11,hoyos12}, which gives the fraction of signal contained in the residual image of the galaxy that cannot be explained by the background fluctuations. It is defined as: 
   \begin{equation}
    RFF = \frac{\sum_{i,j}|I_{i,j}-I_{i,j}^{\rm GALFIT}| - 0.8\times\sum_{i,j}\sigma_{i,j}^{\rm bkg}}{\sum_{i,j}I_{i,j}^{\rm GALFIT}},
   \end{equation}
   where $I_{i,j}$ represents the flux at pixel ($i$,$j$) in the
   galaxy image, while $I^{\rm GALFIT}$ is the model created by {\sc
     galfit}. The $rms$ of the background is $\sigma_{i,j}^{\rm bkg}$.
   As discussed previously, we use the same galaxy residual images for
   computation of $RFF$ over the Kron aperture. The factor of 0.8
   enables the expectation value of $RFF$ for purely Gaussian noise
   error image of constant variance to be 0.0.

\begin{figure}
 \includegraphics[width=0.5\textwidth]{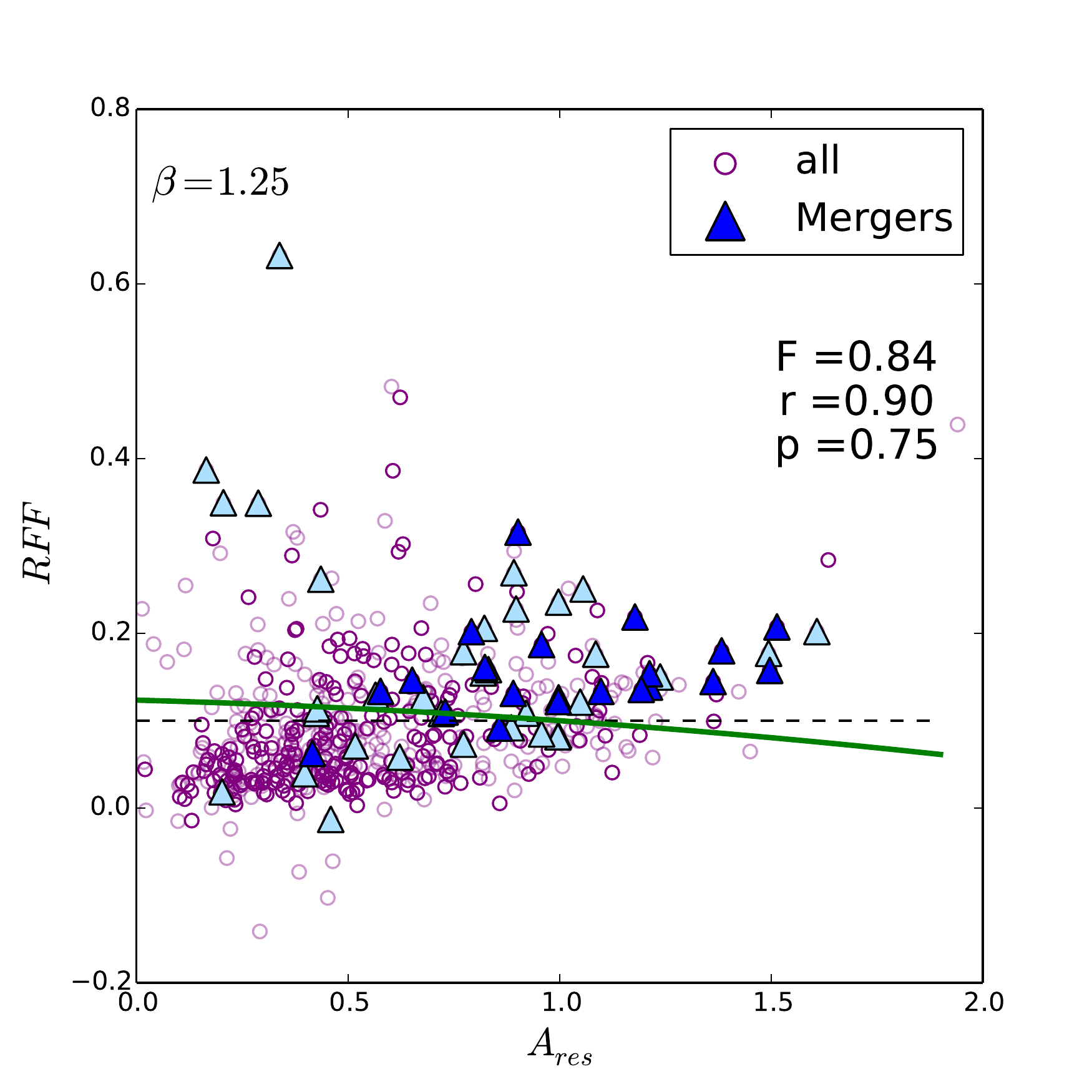}
 \caption{\label{bb}The $RFF-A_{res}$ plane with the statistical best
   border (green) for galaxies classified as mergers, constructed
   using a mass-complete sample. The lighter coloured triangles and circles denote low-mass galaxies excluded from the
   analysis. The dashed line indicates the initial approximation used
   to compute the border. The {\em r} and {\em p} values illustrate
   the completeness and purity of the division: $90\%$ of mergers from
   the training sample lie above the border (true positives) while
   $75\%$ of non-mergers lie below the border (true negatives).  Note
   that the final border remains nearly horizontal, indicating that
   $RFF$ alone provides a good indication of merger status for our
   sample.}
\end{figure}

\subsection[]{Defining galaxy structure}
\label{galst}

\citet{hoyos12} show that $A_{\rm res}$ and $RFF$ are capable of
automatically detecting structural disturbances in galaxies when used
together. Using a training sample of visually identified galaxy
mergers from the low redshift STAGES field \citep{gray09},
\citet{hoyos12} show that these mergers occupy a specific region on
the $RFF$ vs $A_{\rm res}$ plane. This enables a statistical division
of the parent sample into two sub-populations: one containing mergers with some
contamination by non-merging galaxies, and the other almost devoid of
any merging galaxies (a powerful null test).  

We use the same technique on our mass-complete sample to
identify a subsample of structurally disturbed galaxies that we can compare
with our qualitative identifications.  We examine the $RFF$ vs
$A_{\rm res}$ distribution for the entire population of
galaxies in our sample of morphologically classified galaxies with spectroscopic information. We divide our sample by defining a
separating border as a second order polynomial of $RFF$ in terms of $A_{\rm res}$ and separates visually identified mergers from
non-mergers. 

The statistical quality of the two populations is determined by the
$F$-score, $F_{\beta}$ \citep{rijsbergen79} given by
\begin{equation}
    F_{\beta} = \frac{\left(1+\beta^{2}\right)\times p\times r}{\left(\beta^{2} \times p + r\right)},
\end{equation}
where `{\em r}' denotes the sensitivity or completeness of the method,
and `{\em p}' denote specificity or the true negative rate.  The
factor $\beta$ is a control parameter which determines the relative
importance of {\em r} and {\em p}. In this work, we have used
$\beta=1.25$, to be consistent with \citet{hoyos12}.  This border is
then optimized such that it maximizes the F-score. 

In order to apply the $F$-score maximization for detection of galaxy
structure, we use a training sample of galaxies classified visually as
mergers from our parent sample to calculate {\em r} and {\em p}. Figure \ref{bb} shows both populations for the mass-complete
sample, with the separating border represented by the green solid
line.  The galaxies above the solid green line denote the positive
detections of galaxies being mergers, with the merger training sample
retrieved with a high completeness ($\sim90\%$) and a contamination of
$\sim25\%$ by galaxies not classified as mergers. Refer to
\citet{hoyos12}, for the detailed method. 

We note that including low-mass galaxies (log
$M_{*}/M_{\odot} < 10.6$) does not change the border significantly. We
are able to separate the merger subsample from the parent sample,
albeit with lower completeness. For comparison, the galaxies with
masses below the mass completeness are overplotted in light blue in
Figure \ref{bb}. It is clear that this method
gives a clean sample of non-merging galaxies. Comparing to
\citet{hoyos12}, we see a significant flattening of the border
separating the mergers. This can be attributed to the lower S/N of
galaxies at intermediate redshifts, as compared to the local sample of
STAGES galaxies used in \citet{hoyos12}. This is also reflected in
the range of $RFF$ and $A_{\rm res}$ values. However, the important
outcome of this analysis is that while it is possible to separate
regular galaxies from the disturbed galaxies using these two
non-parametric measures, it is the $RFF$ that is the most significant
discriminator of galaxy structure in our sample. Thus, $RFF$ gives a
measure of `roughness' in galaxy structure.

\begin{figure*}
 \hspace{-1.7cm}
 \includegraphics[width=1.1\textwidth]{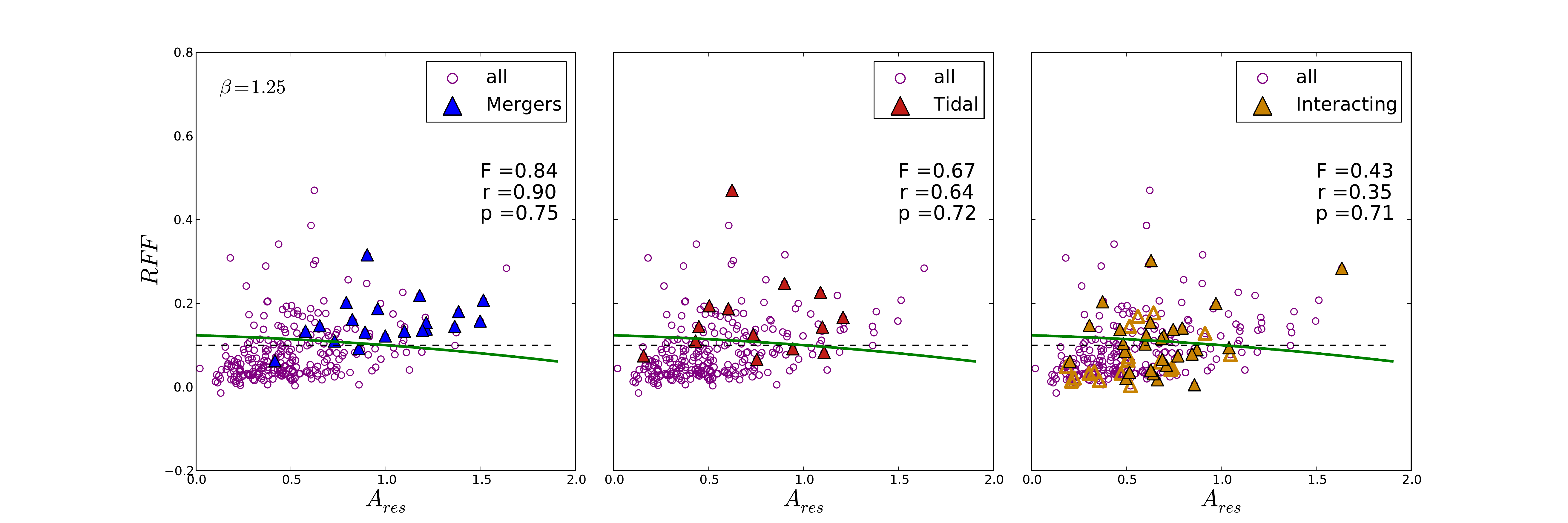}
 \caption{\label{mti} Comparison of the $RFF-A_{res}$ plane for all
   externally disturbed classes of galaxies: mergers (left), tidal
   (centre), and interacting (right), using the mass-complete sample.
   As in Figure \ref{bb}, the dashed and solid lines indicate the
   initial approximation and the final border computed using the
   merger class of galaxies (blue). The $F$-score, {\em r} and {\em p}
   values for each class of galaxies are computed using this
   border. The open triangles in the right panel denote the subclass
   of interacting galaxies nevertheless classified as `symmetric'
   according to our classification scheme (discussed in
   \S\ref{galst}).}
\end{figure*}

Comparing the border determined using the visually classified mergers
to the distribution of galaxies showing disturbances due to other
external causes, we see that this technique is consistent in
separating structurally disturbed galaxies. Figure \ref{mti} shows the
merging, tidal and interacting galaxies on the $RFF$ vs $A_ {res}$
plane. The $F$-score, {\em r} and {\em p} for the tidal and
interacting galaxies is computed using the border determined for the
merger training subset. Although the location is comparable for
merging and tidal galaxies on this plane, we find interacting galaxies
extend below the seprating line in $RFF$. Therefore, in accordance with the
classification scheme, we separate the `true' interacting galaxies
from the visually symmetric interacting galaxies (open yellow triangles),
despite the presence of a companion.  We note that, as expected, the
symmetric interacting galaxies (dry merger candidates, or `0$\&$i')
are the objects populating the region below the separating border in
the $RFF$ vs $A_{\rm res}$ plane.

\section[]{Link between the qualitative and quantitative structure}
\label{sec:VQstruct}

\begin{figure*}
 \includegraphics[width=1\textwidth,left]{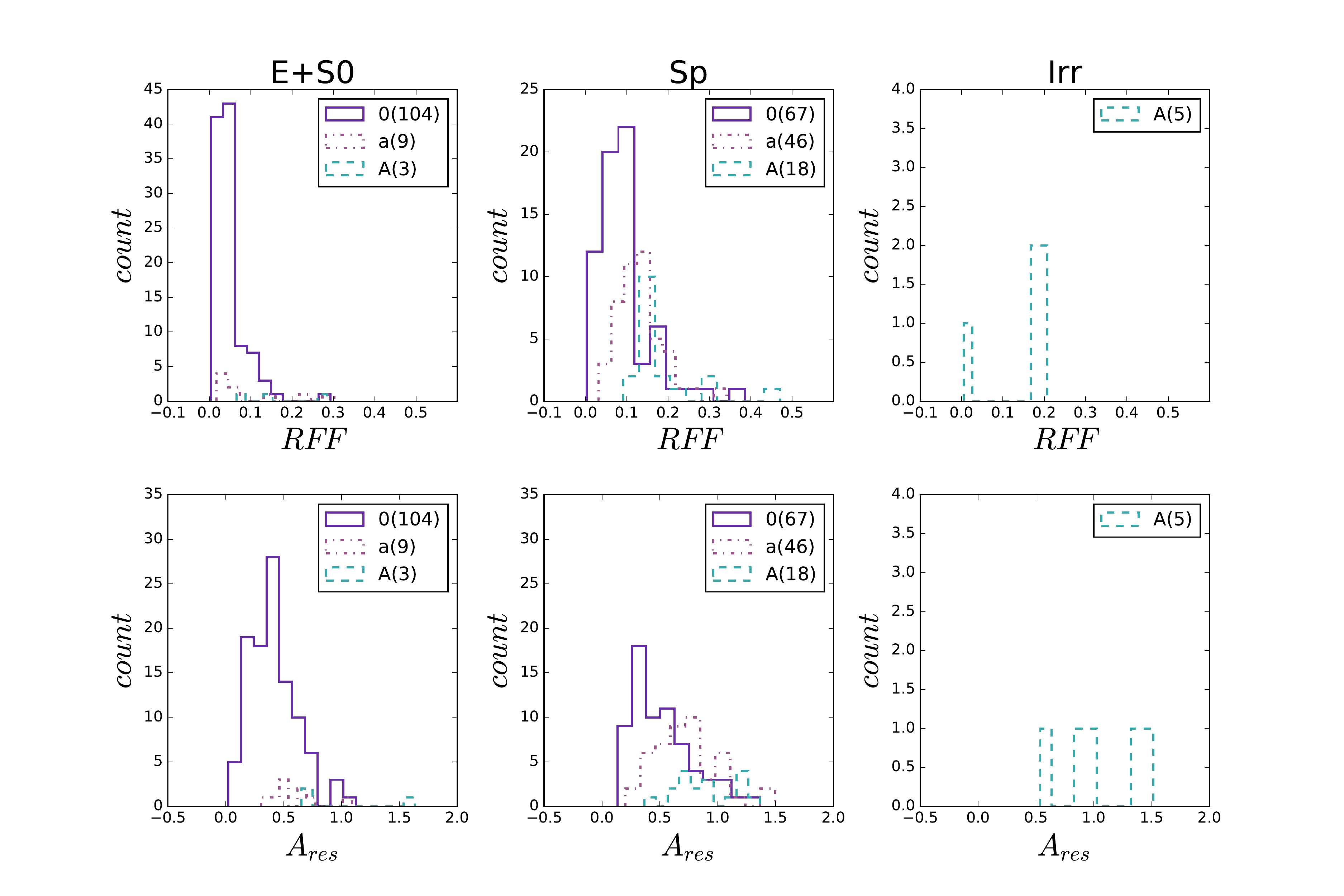}
 \caption{\label{qualqas} Separating quantitative measurements of
   structure by visual asymmetry.  The histograms show the
   distributions of $RFF$ and $A_{\rm res}$ for visually symmetric
   (0), mild (a) and strongly asymmetric (A) galaxies at fixed
   morphology. Early-type galaxies (E+S0) show little quantitative or
   qualitative evidence for asymmetry or disturbance.  Both top and
   bottom panels for spiral galaxies (middle) clearly show a
   separation in $RFF$ and $A_{\rm res}$, with higher values
   corresponding to the strongest visual asymmetries.}
\end{figure*}

We next connect the quantitative measures of galaxy structure with our
visual classification scheme, recalling the need to control for
morphology. Using the CAS system, \citet{conselice03} evaluate the
relative variation in the `Asymmetry' of galaxies when we consider
their morphologies. Galaxies with early-type morphologies displayed
lower asymmetries compared to late-type, star forming or disky
galaxies. We do find consistent results when comparing $A_{\rm res}$
of galaxies with mild/strong visual asymmetry for a fixed morphology
(Figure \ref{qualqas}). We see that galaxies with
elliptical/lenticular morphology are generally visually symmetric
(`0') with very low structural asymmetry in their residuals. The
spiral galaxies, however, show a distinct separation in structural
asymmetry in the residuals for mildly and strongly visually asymmetric
galaxies, confirming that visual and structural asymmetry are strongly
correlated. The irregular galaxy sample is small, but as expected they are all asymmetric with typically high $RFF$ and $A_{\rm res}$ values.

\begin{figure}
 \hspace{-0.9cm}
 \includegraphics[width=0.95\textwidth]{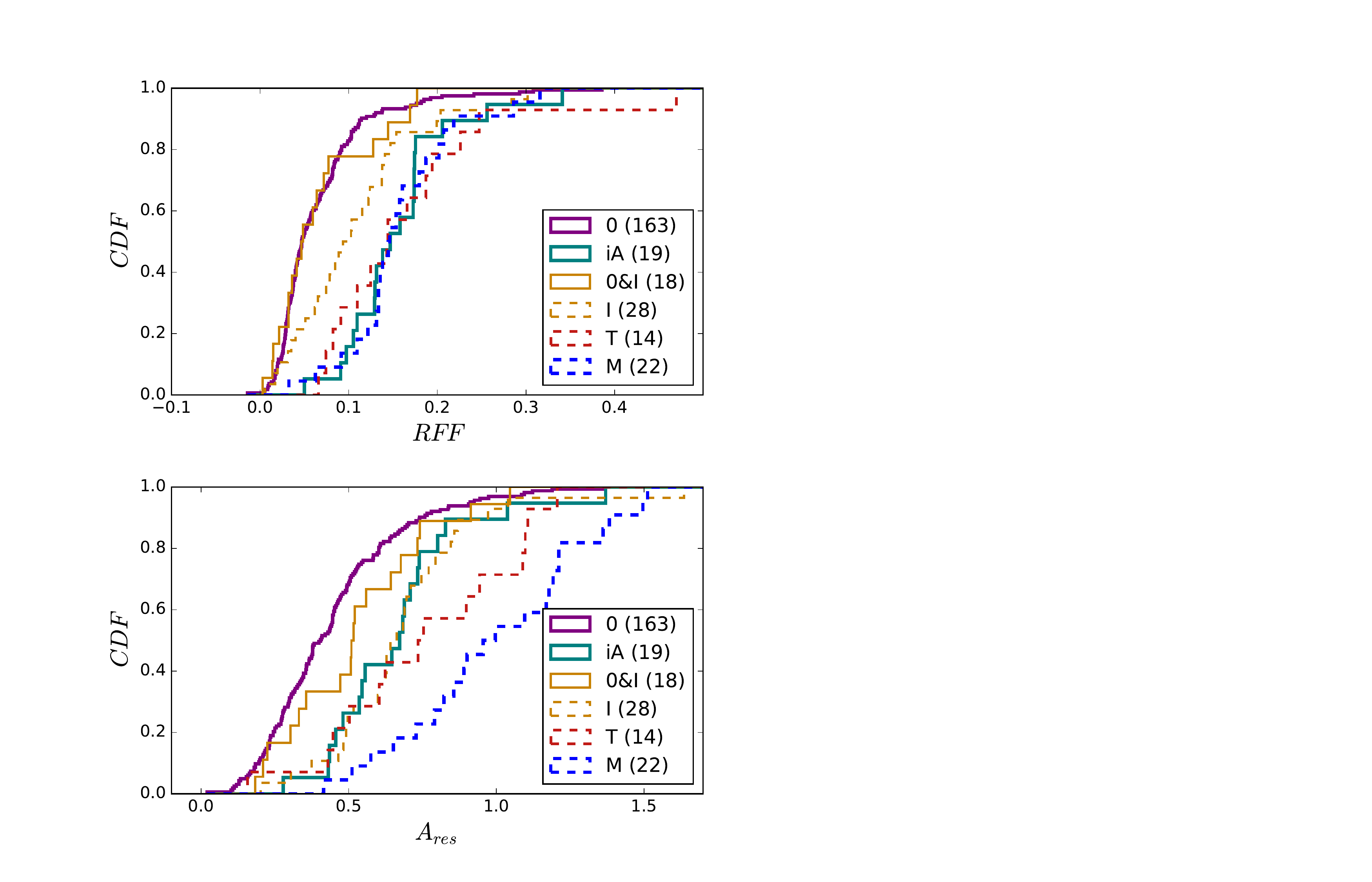}
 \caption{\label{strff} Distributions of two quantitative measurements
   of structure by visual disturbance class.  Here we show the
   cumulative distributions of $RFF$ (top) and $A_{\rm res}$ (bottom)
   for galaxies separated according to the degree and source of any
   visual disturbance.  The distributions are labelled according the
   physical processes introduced in \S\ref{subsecvisdist} and Table
   \ref{table}. However, for the interacting galaxies, we have plotted
   the separate distributions for the subclass of symmetric
   interacting galaxies (`0$\&$I') whereas `I' denotes the main class
   of interacting galaxies displaying obvious visual asymmetry.  Note
   that while $RFF$ clearly separates symmetrical galaxies (`0',
   `0$\&$I') from the remaining classes, $A_{\rm res}$ has further
   discriminatory power according to the \emph{cause} of any
   asymmetry, with internal asymmetries (`iA') and interacting
   galaxies (`I') having some of the lowest values and mergers (`M')
   having the largest. }
\end{figure}

We next consider the variation of quantitative measures of structure with the visually determined causes of disruption.
Figure \ref{strff} shows the distribution of $RFF$ and $A_{\rm res}$
for galaxies disrupted due to different mechanisms. If we consider
only the $RFF$ of galaxies, we find that $RFF$ alone is not able to
distinguish the tidal, merging and internally asymmetric galaxies,
indicating that $RFF$ is sensitive to the degree of disturbance rather
than the cause of the disturbance. This result is
also graphically demonstrated when comparing the best border for
galaxies in different disturbance classes in Figure
\ref{mti}. Additionally, the subclass of
symmetric interacting galaxies (`0$\&$I') seem to have $RFF$
distributions similar to galaxies with regular morphologies, although
the distribution for the true interacting galaxies lies in between.

The lower panel of Figure~\ref{strff} shows the distribution of $A_{\rm res}$ for different disturbance classes. Interestingly, a significant stratification is seen in the distribution of $A_{\rm res}$ for different disturbance classes, with undisturbed galaxies having a low $A_{\rm res}$ and mergers showing extreme values of $A_{\rm res}$.

We conclude that $RFF$ is able to separate galaxies with disturbed structure from those with regular undisturbed structure, but has little discriminatory power to differentiate between the different types (or causes) of such disturbances. On the other hand, $A_{\rm res}$ is more sensitive to the different types (or causes) of structural disturbance in the galaxies. In simple terms, $RFF$ can be used as a measure of the degree of structural disturbance, while $A_{\rm res}$ provides information on the cause of it. Or, more precisely, a combination of both parameters can be used to provide information on both the degree and the cause of galaxy deviations from symmetry.

\section[]{Structure and star formation vs global environment}
\label{sec:structSF}

\subsection[]{Effect of global environment on galaxy disturbances}
\label{dmorph}        

This analysis uses the spectroscopic sample with visual classifications to compare the properties of galaxies as a function of global environment (e.g., cluster vs. field) rather than a more continuous measure of local environment. That will be the object of a subsequent paper (Kelkar et al. 2017, in prep).

In Section~\ref{subsecenv} we described a full redshift-controlled field sample together with a mass-complete subsample. The full sample has the advantage of being significantly larger, but it suffers from incompleteness for galaxies with log $M_{*}/M_{\odot} <$ 10.6. Nevertheless, because the selection and the observation of cluster and field galaxies over the relevant redshift range is identical, the incompleteness should affect field and cluster galaxies equally. With this in mind, when carrying out comparisons between the properties of cluster and field galaxies it should be safe to use the full redshift-controlled sample. Nevertheless, we will carry out a parallel analysis using the smaller mass-complete subsample to test whether our conclusions depend on the exact sample that we use. In general, we find that the conclusions described below for the full sample are consistent with the ones we obtain for the mass-complete sample within the statistical uncertainties. Further, to remove additional effects brought in by the fact that galaxy morphology depends strongly on environment \citep{dressler84,treu03,desai07}, we look at the disturbance content of galaxies in clusters and field at fixed morphology (Table~\ref{table}). We find that the fractions of galaxies classified visually as interacting, tidal and merging do not seem to depend on the environment.

\subsection[]{Distribution of galaxy disturbances and star formation as a function of global environment}
\label{}

As discussed in the introduction, the internal and external physical
mechanisms in various galaxy environments are responsible for the
transformation of galaxy structure as well as star formation. Although
disentangling the relative importance of these processes is difficult,
quenching in the star formation of galaxies is observed in dense
environments \citep{balogh07,haines13,kovac14}. With the aim of studying the possible
links between the quenching of star formation and the morphological
change in galaxies, we next look at the star formation properties of
structurally disturbed galaxies. 

\begin{figure}
 \includegraphics[width=0.5\textwidth]{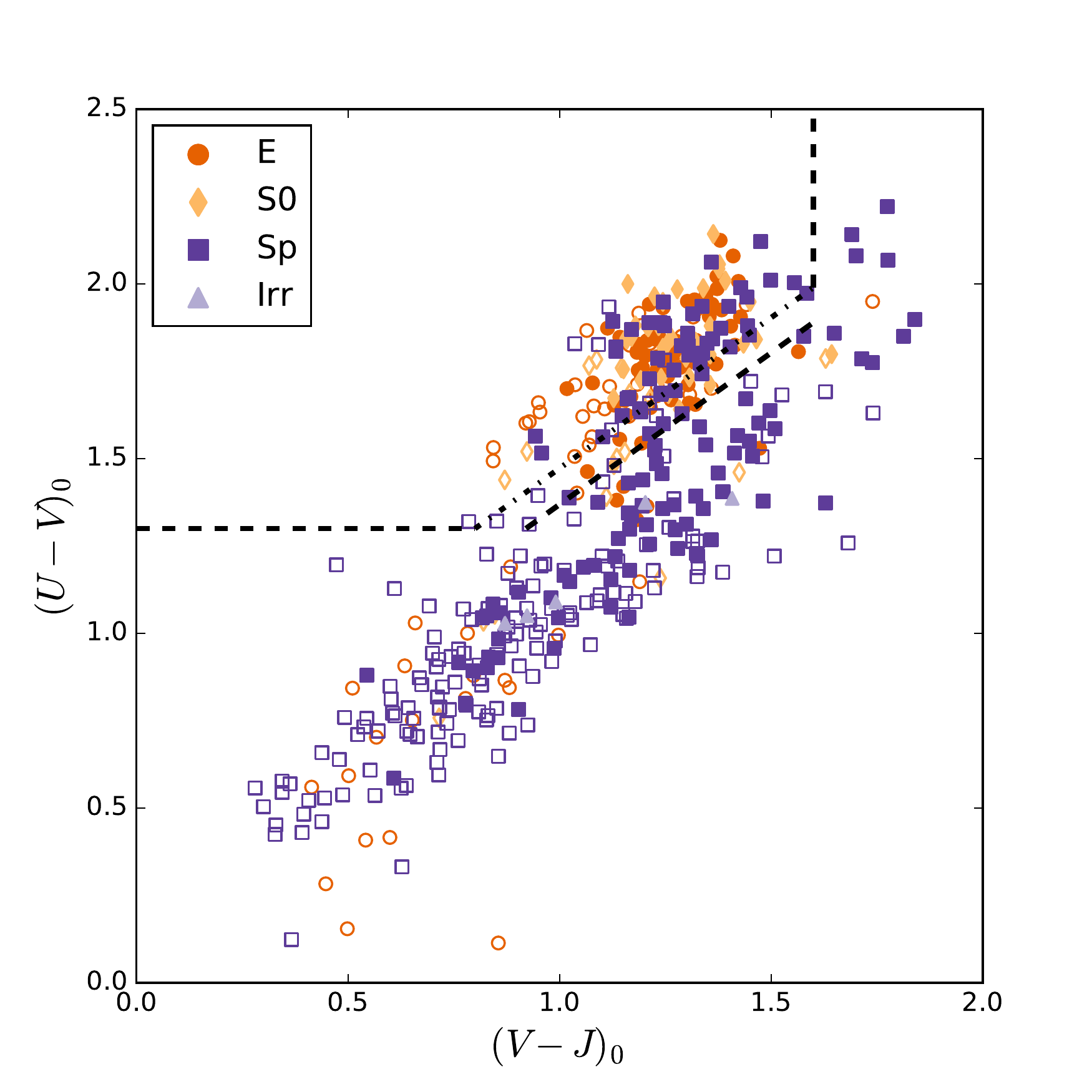}
 \caption{\label{uvj} Rest-frame $UVJ$ colour plot for the ellipticals (circle/orange), lenticulars (diamond/yellow), spirals (squares/violet), and irregulars (triangles/light violet). The galaxies are defined as passive 
   and star-forming according to the boundaries of 
   \citet{williams09} for the two redshift bins: $0.5<z<1$ (dashdot) and $z>1$ (dashed). The empty symbols denote galaxies with
   masses below the mass completeness limit.}
\end{figure}

It was found by \citet{wuyts07} and \citet{williams09} that galaxies show a strong bimodality on the rest-frame $(U-V)$ vs $(V-J)$ colour-colour space, with the actively star-forming galaxies following a diagonal path and the quiescent galaxies populating upper-left region on this space \citep[see also][]{wolf05,labbe05}. Moreover, the $(U-V)$ vs $(V-J)$ plane is more robust to separate the dusty star-forming galaxies from the passive galaxies, as compared to the single colour selection.  
Therefore, we construct a rest-frame $(U-V)$ vs $(V-J)$ colour plot ($UVJ$ hereafter) to
distinguish the passive and star-forming population \citep[Figure \ref{uvj}; see also][]{patel12}. The $UVJ$ plot shows that the low-mass galaxies (log $M_{*}/M_{\odot} \leqslant 10.6$) are bluer in colour,
as expected from the existing correlations between mass, metallicity, star-formation rate and dust extinction \citep{llopez10}.

The empirical selection criteria for passive galaxies, as introduced by \citet{williams09}, highlights the observed bimodal distribution of galaxies on the $UVJ$ plane at low-$z$, and the subsequent weakening at high-$z$. Figure~\ref{uvj} shows such a distribution for galaxies in our sample with different morphologies. The boundary separating star-forming and passive galaxies in the $UVJ$ plane is somewhat arbitrary, and \citet{williams09} found that this boundary is weakly redshift dependent. It will also depend on the exact photometric bands used in the observations. In Figure~\ref{uvj} we show the boundaries selected by \citet{williams09} for two redshift ranges, $0.5 < z < 1$ and $z > 1$. Given the redshift of our galaxies, the $0.5 < z < 1$ boundary should be, in principle, more appropriate. However, we notice that the $z>1$ boundary seems to do a much better job at separating the bimodal colour distribution than the lower redshift one, in particular if we take into account the location of galaxies with different morphologies. A density-mapping analysis corroborates this visual impression, revealing that galaxies with early-type morphologies populate the upper left region of the $UVJ$ plot, as defined by the $z>1$ boundary, while the late-type spiral and irregular galaxies occupy the diagonal sequence, and the separation is significantly better with this boundary than with the $0.5 < z < 1$ one. We concluded that for our specific dataset, the $z>1$ boundary is better at separating star-forming from non-star-forming galaxies. Explicitly, we use the boundary defined by

\begin{numcases}{}
    & $(U-V)\geqslant1.3$ \label{positive}
   \\
    & $(U-V)\geqslant0.88\ast(V-J)+0.49$ \label{negative}
   \\
    & $(V-J)\leqslant1.6$ \label{negative2}
\end{numcases}
to separate passive non-star-forming galaxies (inside the upper-left box) from star-forming ones.

\begin{figure*}
 \centering
 \includegraphics[width=1\textwidth]{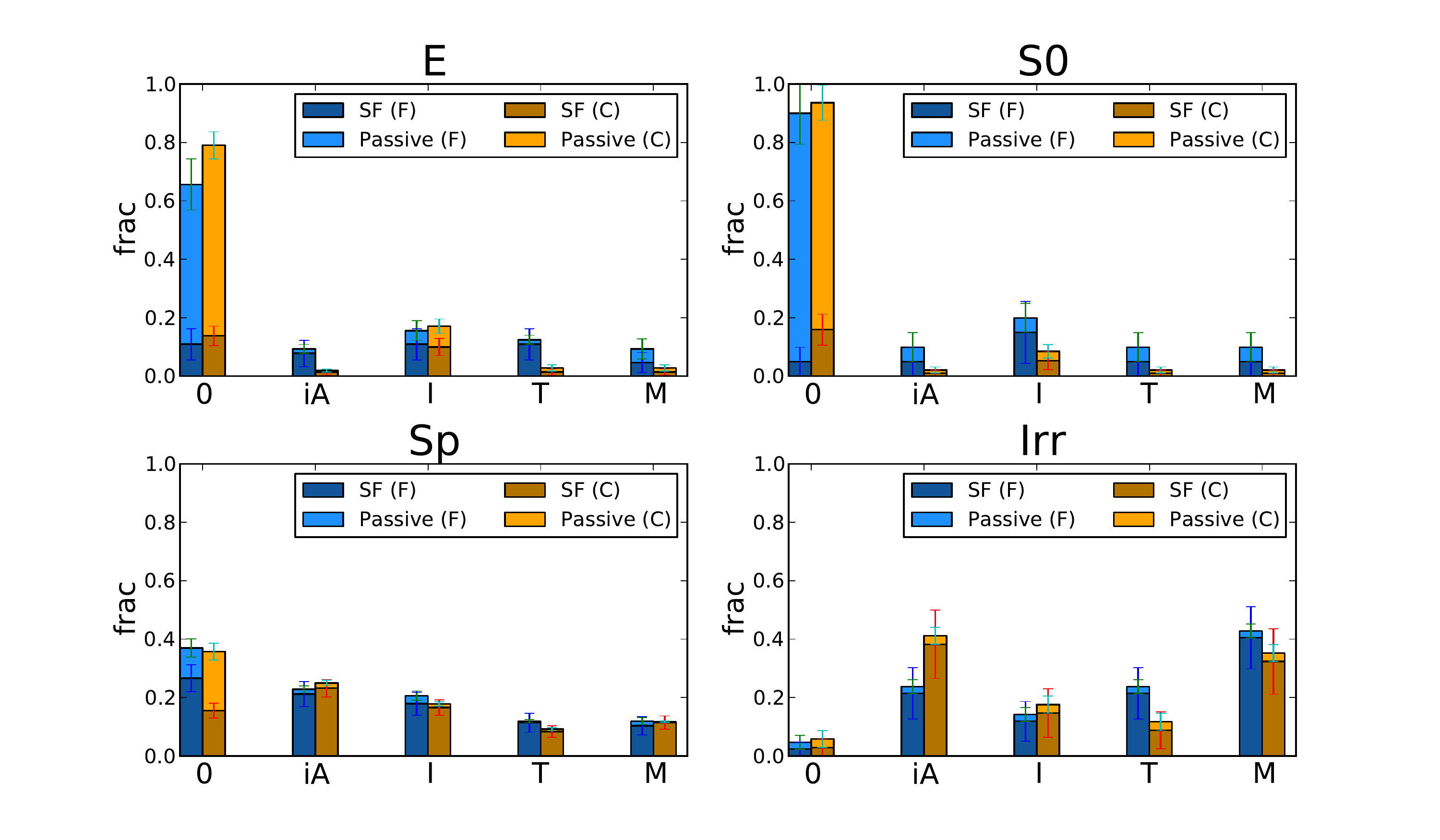}
 \caption{\label{stdistmorph} Visual galaxy structure at fixed
   morphology according to the global environment. Each panel shows the
   fractions of galaxies in different disturbance classes for a given
   morphology, in the field (blue/left column) and cluster (orange/right column) environments.  The
   dark shaded region in each bar shows the fraction of star-forming
   galaxies in that class of morphology, disturbance class, and
   environment.  In both cluster and field environments, less than
   half of all spiral galaxies are visually symmetric (`0'), while for
   asymmetric galaxies the causes of the disturbance are roughly
   equally distributed between internal causes (`iA'), interactions (`I'), tidal forces
     (`T'), and mergers (`M').  Note also that for symmetric spirals,
     a marginally significantly higher fraction ($2.3\sigma$) in cluster environments are
     passive than in the field. Conversely, passive spirals are only
     found in the visually symmetric class.}
\end{figure*}

Figure \ref{stdistmorph} shows the fraction of passive and star
forming galaxies of each morphology type in various structural
disturbance classes, and global environments. It is evident from each
of the four panels that galaxies with early-type morphology are
visually smoother compared to the galaxies of late-type/irregular
morphology, which is expected from the distinguishing physical
properties of these two subpopulations. 

Most spirals show asymmetries, and for those, a flat
distribution of spirals across the disturbance classes (most of which
are external in origin) is observed. Furthermore, in each disturbance class
nearly all are star-forming, indicating a strong correlation of star
formation with external causes of structural disturbances. We also note a relatively higher fraction of cluster star-forming spirals displaying tidal features as compared to the field. These spirals could be the potential candidates with `jellyfish' morphology \citep{ebeling14,poggianti16}. Likewise, all of the irregulars are both disturbed
according to their stellar distribution, and star-forming according to
their photometry. We can, thus, say that external processes lead to star formation in galaxies irrespective of their morphology. However, the star formation can also be triggered internally in galaxies.
In general, it is seen that the passive population in the sample --
irrespective of their morphology -- tend to have lower $RFF$ and lower
$A_{\rm res}$, and hence are structurally smoother and symmetric. This
correlation, however, seems to be independent of the global
environment.

\subsection[]{A population of smooth passive spirals in clusters}
\label{passsp} 

We make particular note of those spiral galaxies that are
simultaneously classified as visually smooth/undisturbed and
passive. Figure~\ref{stdistmorph} shows that most are found to reside
in clusters, with the majority ($>70\%$) of these having stellar masses greater
than the mass completeness limit, both in cluster and field
environment. This observation agrees with the findings from works such
\citet{poggianti99, wolf09, cantale16}, Rodriguez del Pi\~{n}o et al. (2016, MNRAS, in press), who both find a significant fraction of
passive spirals in the cluster environment that may represent a key
transition population undergoing slow environmental quenching \citep[See also][]{bamford09, masters10}. Most
recently, \citet{hoyos15} reported that optically passive spiral
galaxies in clusters, at a given mass, tend to have lower star
formation rates and smoother structure as compared to the galaxies
in field. This result is particularly relevant here because these authors used quantitative structural measurements similar to the ones we present in this paper.

\begin{figure*}
 \includegraphics[width=1\textwidth]{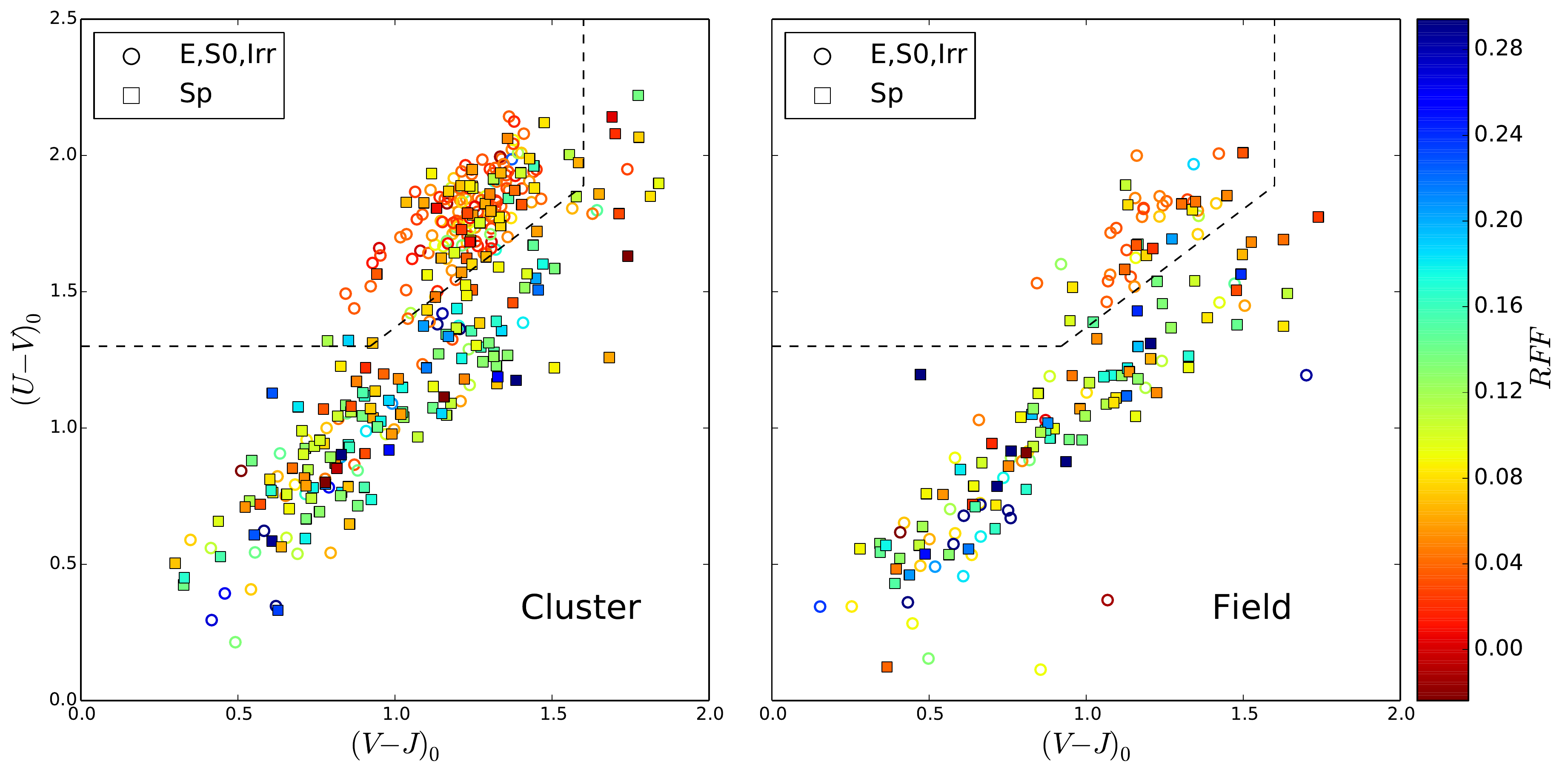}
 \caption{\label{uvjr} The $UVJ$ plot is colour-coded according to
   $RFF$ values for the cluster (left panel) and field (right panel)
   galaxies: redder colours indicate smoother, undisturbed galaxies
   while bluer colours denote `roughness', irrespective of galaxy morphology or global environment. As in Figure \ref{uvj},
   the dashed lines show the selection boundaries for passive
   galaxies. Complementing Figure~\ref{stdistmorph}, both panels show that
the passive undisturbed spirals have lower RFF, indicating a
smoother structure. }

\end{figure*}

\begin{figure*}
 \includegraphics[width=1\textwidth]{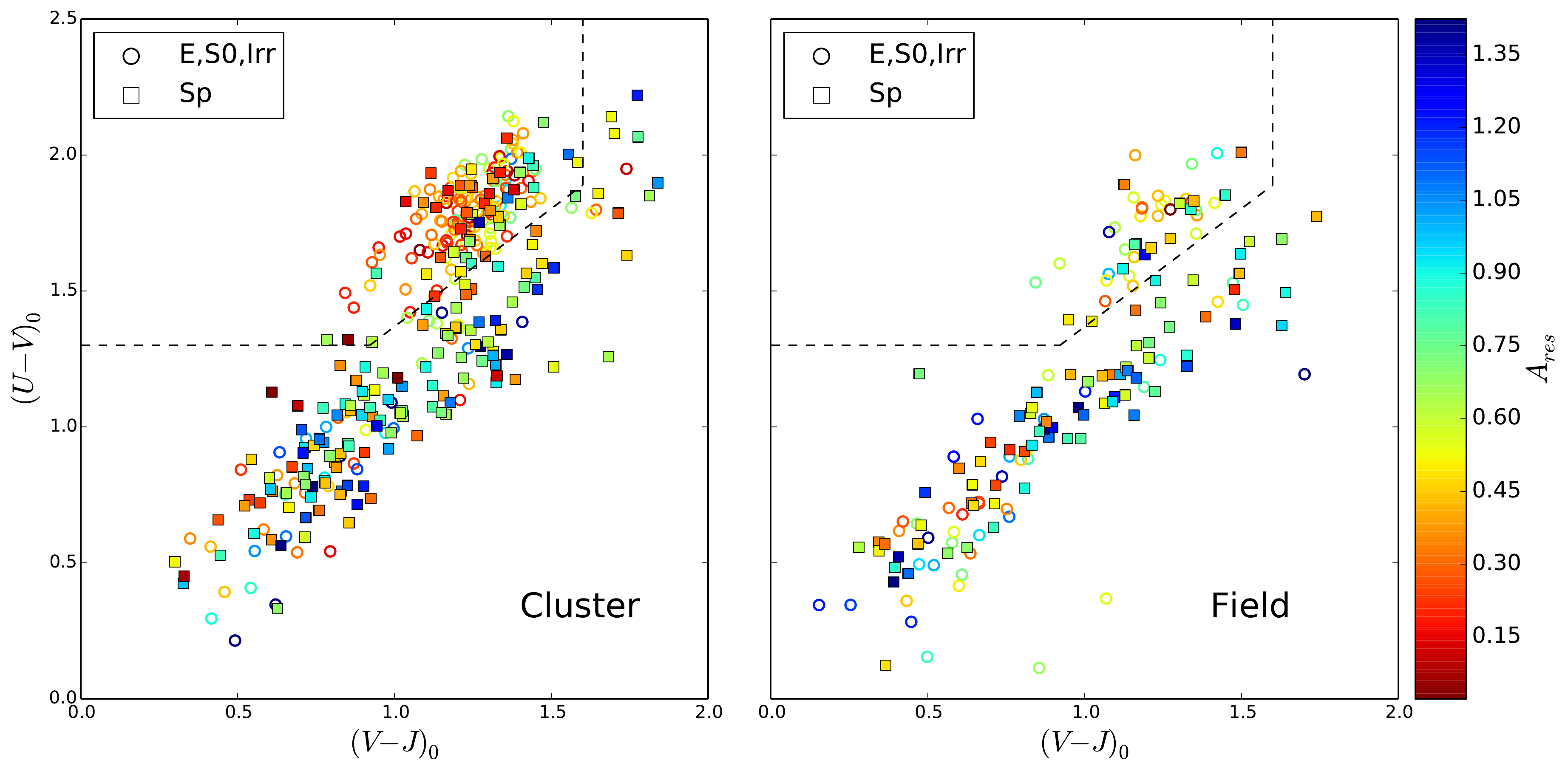}
 \caption{\label{uvja} Similar to Figure \ref{uvjr}, the $UVJ$ plot is
   here colour-coded according to $A_{\rm res}$ for cluster (left
   panel) and field (right panel) galaxies. Star-forming galaxies are
   consistently found to be more quantitatively asymmetric than
   passive galaxies irrespective of morphology or global environment. Moreover, complementing Figure~\ref{uvjr}, the  passive spirals in clusters are found to be more symmetric with low $A_{\rm res}$.}
\end{figure*}

To test whether quantitative measurements of galaxy disturbance support the findings based on our visual diagnostics, we present in Figures~\ref{uvjr} and ~\ref{uvja} the rest-frame $UVJ$ diagram colour-coded with respect to $RFF$ and $A_{\rm res}$, for cluster and field galaxies. Complementing Figure~\ref{stdistmorph}, both panels in Figure \ref{uvjr} show that these passive undisturbed spirals have lower $RFF$, indicating a smoother structure.  
This is further enhanced in Figure
\ref{uvja}, where we see that passive spirals in clusters are much
more symmetric with low $A_{\rm res}$. This observation combines the result from Figures \ref{strff} and
\ref{stdistmorph} demonstrating the external nature of structural
disturbances for the majority of the asymmetric spirals, and the different
behaviour of $A_{\rm res}$ in the different disturbance classes.

\begin{figure}

 \includegraphics[width=0.5\textwidth]{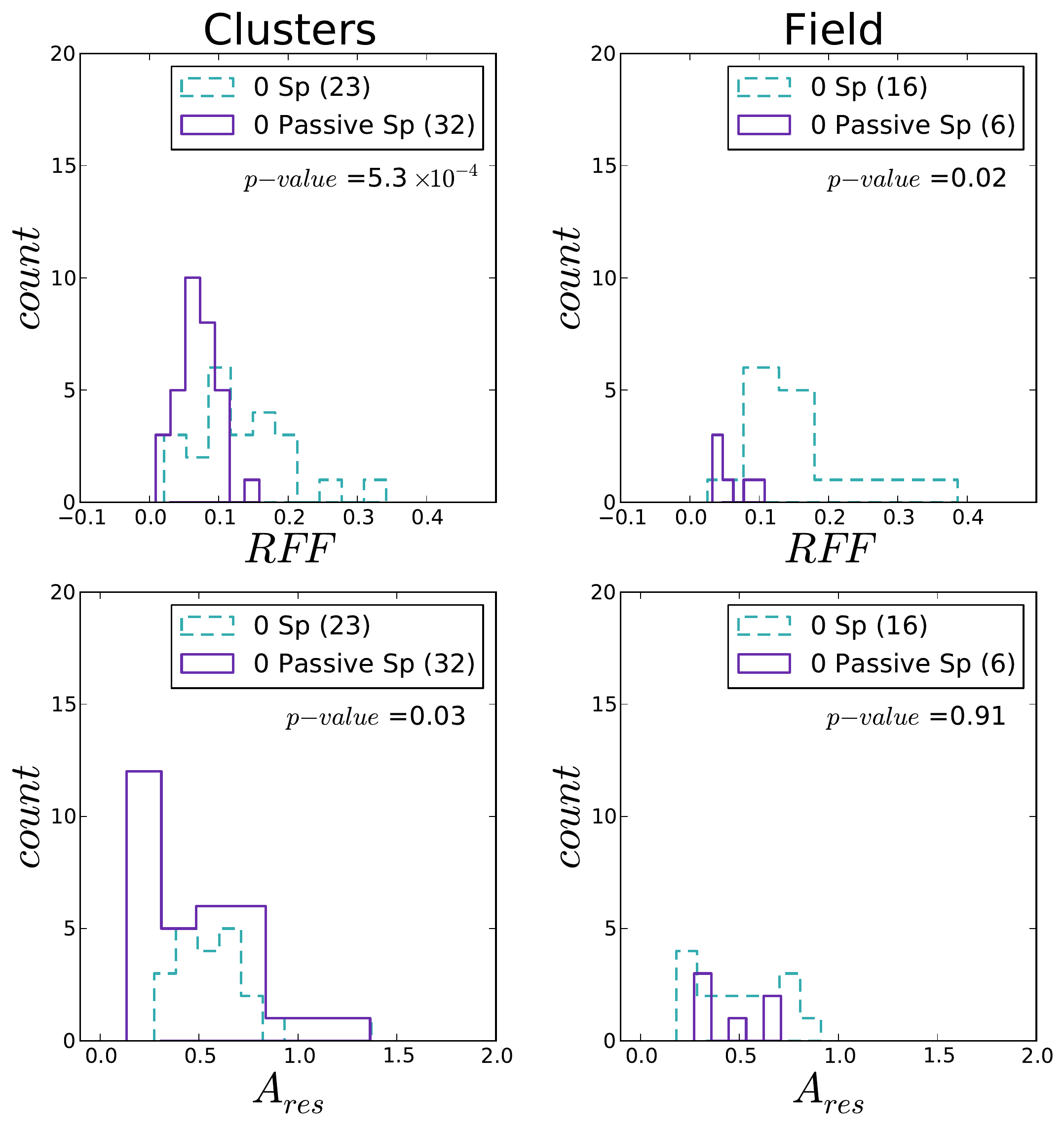}
 \caption{\label{rff_ps} $RFF$ and $A_{\rm res}$ distributions for
   spiral galaxies in cluster (left panel) and field (right panel) environments. The solid purple histograms
   show the subset of passive symmetric spirals, as discussed in \S\ref{passsp}. The inset `$p-$values' give the probability from two sample K--S test,
   showing that in addition to being visually symmetric, the cluster passive spirals are
   quantitatively `smoother'.}
\end{figure}

We use two-sample Kolmogorov–Smirnov (K--S) tests to check whether the $RFF$ and $A_{\rm res}$ distributions for these passive spirals, and the regular undisturbed spirals are statistically similar. Figure~\ref{rff_ps} compares the $RFF$ and $A_{\rm res}$ distributions for passive spirals and regular undisturbed spirals, both in cluster and field environments. The K–S tests yields probabilities of $5.3\times10^{-4}$ and $0.02$ for these distributions to be the same for cluster and field galaxies respectively. This emphasises the fact that passive spirals tend to show statistically smaller $RFF$ values –and are therefore smoother than star-forming ones irrespective of their global environment. However, the distributions of $A_{\rm res}$ for passive and star-forming spirals appear to be only marginally different in clusters (K--S test probability of $0.03$), while the small number statistics prevent a robust comparison for field spirals.

These results reinforce out findings from Figure~\ref{stdistmorph}, implying that the effect of the cluster environment on the spiral galaxy population is to increase the fraction of passive smooth spiral galaxies without destroying their spiral morphology.
This would signify that spirals on entering clusters become structurally smooth due to the quenching of their star formation followed later by morphological transformation, perhaps into S0s. This implies that that the mechanisms ultimately responsible for the quenching of these galaxies’ star formation in clusters must be reasonably gentle, affecting primarily the gas while leaving the galaxies' stellar structure largely unchanged. These galaxies become smoother due to the suppression of the star formation itself, since `rough' structures such as HII regions would disappear \citep[see, e.g.,][]{hoyos15}. Gas-driven mechanisms such as ram-pressure striping are therefore strongly favoured. These conclusions are in good agreement with the findings of \citet{bosch13} based on observation of the lower redshift STAGES field \citep{gray09}, which show that red spirals display distinct asymmetries in their gas rotation curves, and are therefore preferentially experiencing ram-pressure stripping, as compared to normal spiral galaxies.

Complementary conclusions were obtained by \citet[][see their Figure~10]{cantale16} using the $UVJ$ colours of disks in the EDiSCS dataset. These authors find that $\sim50\%$ of cluster spirals have redder disks than their field counterparts at fixed morphology, but they also find evidence that spiral galaxies must have continued forming stars for a significant period of time after their accretion into the clusters, getting quenched thereafter on a timescale of a few Gyrs.

\subsection[]{A small population of star-forming cluster S0s}

Turning our attention to lenticular (S0) galaxies, Figure~\ref{stdistmorph} indicates that, as expected, the vast majority of these galaxies are symmetric and passive both in clusters and in the field. However, although the numbers are small and the statistical uncertainties very large, there seems to be some marginal evidence suggesting the presence of an excess of star-forming S0 galaxies in
clusters with respect to the field. Some of these star-forming S0s are asymmetric, showing signs of perturbation (interactions, mergers and tidal features), but there seems to be also a population of symmetric star-forming S0s in clusters which is absent in the field. Specifically, we do not find a single symmetric undisturbed star-forming S0 in the field, although the expectation value of their fraction shown in Figure~\ref{stdistmorph} is not $0$ (cf. Equation~\ref{Equation: Frequency error}). Although the undisturbed lenticulars have a wide spectrum of stellar masses in all environments, we note that the majority ($> 80$\%) of the symmetric star-forming cluster lenticulars have relatively low stellar masses that are below the completeness limit. It is therefore not impossible, albeit unlikely, that these galaxies may have been missed preferentially in the field. If that is not the case, this result supports the findings of \citet{johnston14}, suggesting that in the putative transformation of spirals into S0s in clusters, a final episode of star formation takes place in the central regions (bulges) of these galaxies after the disk star formation has ceased. In this scenario, the gas is removed from the disk of the spirals, while some gas remains in (and/or is channelled to) the bulge, where this final gasp of star formation takes place. This process probably requires an external cause and it may therefore be cluster-specific. That could explain why this final episode of star formation is not observed in undisturbed field S0s, where other formation mechanisms may need to be invoked.

\section{Conclusions}
\label{secconclusions}

In this paper we present a detailed analysis of the structure of a sample of field and cluster galaxies at intermediate redshift ($0.4<z<0.8$) using Hubble Space Telescope images from the ESO Distant Cluster Survey (EDisCS) which approximately sample the $B$-band in the rest-frame of the galaxies.  We combine this structural information with extensive photometric and spectroscopic data to study the links between galaxy structure and other internal properties such as mass, morphology and star-formation history, and how these are affected by the global environment where the galaxies live.

We have analysed the galaxies' structure following two parallel methods. In the first one, we visually inspected the galaxies' {\em HST} images and classified them into symmetric and asymmetric; the asymmetric class was further divided into subclasses that try to identify the likely cause of the asymmetry (internal asymmetry, galaxy-galaxy interactions, tidal interactions, and mergers). The second method uses quantitative non-parametric measurements of the galaxies' deviation from a smooth symmetric light distribution. An elliptical S\'ersic model is first fitted to the galaxies' {\em HST} images, and the residuals are then quantified using the $RFF$ (Residual Flux Fraction, measuring the fractional contribution of the residuals to the total galaxy light, taking into account the noise), and $A_{\rm res}$ (the asymmetry of the residual light distribution). The main conclusions of this structural analysis are: 

\begin{itemize}
\item {The qualitative (visual classification) and quantitative ($RFF$ and $A_{\rm res}$) assessments of galaxy structure provide consistent and complementary information.}

\item {$RFF$ is able to separate galaxies with disturbed structure from those with regular undisturbed structure, but has little discriminatory power to differentiate between the different types (or causes) of such disturbances. On the other hand, $A_{\rm res}$ is more sensitive to the different types (or causes) of structural disturbance in the galaxies. A combination of both parameters can therefore be used to provide information on both the degree and the cause of galaxy deviations from symmetry.}

\end{itemize}

We then link this structural information with the galaxies' masses, morphologies and star-formation histories, and conclude that:

\begin{itemize}

\item{As expected, the vast majority of elliptical and S0 galaxies are smooth and symmetric, while all irregular galaxies are ``rough'' and asymmetric. Statistically, spiral galaxies tend to have higher values of $RFF$ and $A_{\rm res}$ than early-type galaxies.}  

\item {Over 60\% of all spiral galaxies are visually classified as showing some degree of asymmetry. Of these, about one third exhibit asymmetry of internal origin (due, e.g., to the presence of large star-forming regions), while the rest show signs of galaxy-galaxy interactions, tidal interactions or mergers in comparable proportions. }

\item{In agreement with the results of \citet{hoyos15}, we find that $RFF$ correlates strongly with the star-formation activity of the galaxies: star-forming galaxies tend to have much ``rougher'' structures. }

\end{itemize} 

Finally, the global environment (cluster vs.\ field) of the galaxies is taken into consideration, and we find that:

\begin{itemize}

\item {At fixed morphology, there are no significant differences in the distribution of the disturbance classes of cluster and field galaxies.}

\item {About $40$\% of all the spiral galaxies are classified as symmetric and undisturbed both in clusters and in the field. However, the fraction of these that are passive (i.e., non-starforming) is twice as large in clusters than in the field: about half of the cluster symmetric spirals are passive, vs. only one quarter in the field (with a significance of $2.3\sigma$). These passive spirals are not only visually symmetric, but also quantitatively smoother (i.e., have smaller $RFF$ values) than star-forming ones.} 

\item{While nearly all lenticular galaxies are visually symmetric and undisturbed both in clusters and in the field, \emph{all} the field ones are passive, while nearly $\sim$20\% in the clusters are star-forming.}

\end{itemize}

These results have significant implications for the evolution of spiral galaxies falling onto clusters and their subsequent transformation. Spirals entering clusters become structurally smooth (and red) due to the quenching of their star formation, but retain their spiral morphology for a while. The morphological evolution follows later, transforming them, probably, into S0s. The mechanism(s) ultimately responsible for the quenching of these galaxies’ star formation in clusters must primarily affect the gas while leaving the galaxies’ stellar structure largely unchanged. Gas-driven mechanisms such as ram-pressure striping (where the disk gas is partially or totally stripped) and/or starvation/strangulation (where the gas supply is truncated), are therefore favoured. These conclusions are in good agreement with the findings of B\"osch et al. (2013) based on observation of the lower redshift STAGES field (Gray et al. 2009), which show that red spirals display distinct asymmetries in their gas rotation curves, and are therefore preferentially experiencing ram-pressure stripping, as compared to normal spiral galaxies. Similar conclusions were obtained by \cite{jaffe11} for EDisCS galaxies. This general scenario also agrees with observations indicating a rapid build-up of red-sequence galaxies earlier than the build up of early-type galaxies as seen in clusters \citep{desai07,delucia07,wolf09,rudnick09,rudnick12,cerulo16}.

At a more speculative level, our analysis also provides some clues on the putative transformation of spirals into S0s. 
The star-forming S0s we find in the clusters (but not the field) could be the descendants of the spiral galaxies experiencing a last episode of star formation before becoming S0s, supporting the findings of \cite{johnston14}. These authors suggest that when spirals transform into S0s in clusters, a final episode of star formation takes place in the central regions (bulges) of these galaxies after the disk star formation has ceased.  In this scenario, the gas is removed from the disk of the spirals, while some gas remains in (and/or is channelled to) the bulge, where this final gasp of star formation takes place. This process probably requires an external cause (e.g., ram pressure) and it may not work in the field. This could explain why this final episode of star formation is not observed in undisturbed field S0s, where other formation mechanisms may need to be invoked. 

Focusing on the general question of ``nature'' vs.\ ``nurture'' in galaxy evolution, it is now clear that the processes leading to the cessation of star formation depend both on internal properties (e.g., stellar mass) and environment, with the dominant quenching mechanisms being environmentally-driven or mass-driven for different mass ranges, cosmic epochs, and environments \citep{ypeng10,thomas10}. Studies at lower \citep{baldry06,wetzel12} and higher redshifts \citep{muzzin12} show that the quiescent fraction is correlated with both stellar mass and environment, and this relationship is maintained even at $z>1$ \citep{quadri12,lizzie16,hatch16}. With the importance of environmental quenching increasing with cosmic time and decreasing with stellar mass, our analysis is particularly relevant because we explore the intermediate mass and redshift regimes, where both stellar mass and environment probably play significant roles in shutting down the star formation. In addition, focusing on differences in the internal galaxy structure at fixed morphology has allowed us to uncover subtle environmental effects that broader-brush studies had missed.

However, the work published here does not provide sufficient details on the possible environmental mechanisms at play because we have only considered global environments such as clusters and the field, disregarding more localised effects. This will be the focus of Kelkar et al. (2017, in prep.) where we use tools like the projected phase-space diagram to constrain the detailed environmental history of the cluster galaxies. Moreover, studying directly the timescales associated with the quenching of star formation will provide very valuable complementary information (Wolf et al.\ 2017, in prep.).

\section*{Acknowledgments}

Based on observations made 
with the NAS/ESA \textit{Hubble Space Telescope}, obtained at the Space Telescope Science Institute, which is operated by the Association of Universities for Research in 
Astronomy, Inc., under NASA contract NAS 5-26555. These observations are associated with proposal 9476. Support for this proposal was provided by NASA through grant
from the Space Telescope Science Institute.

Based on observations obtained at the ESO Very Large Telescope (VLT) as a part of the Large Programme 
166.A-0162

We acknowledge the EDisCS team for providing a unique dataset, and useful comments on the work presented in this paper. We also thank Yara Jaff\'e for valuable discussions regarding the visual classifications. We would also like to thank the referee for useful comments and feedback which helped in making the contents of this paper better.
GHR acknowledges the support of NASA grant {\em HST}-AR-12152.001-A, NSF grants 1211358 and 1517815, the support of an ESO visiting fellowship, and the hospitality of the Max Planck Institute for Astronomy, the Max Planck Institue for Extraterrestrial Physics, and the Hamburg Observatory. GHR also acknowledges the support of a Alexander von Humboldt Foundation Fellowship for experienced researchers. GHR recognizes the support of the International Space Sciences Institute for their workshop support.
 



\appendix

\bsp

\label{lastpage}

\end{document}